\documentclass[nofootinbib,aps,pre,twocolumn,showkeys,superscriptaddress,preprintnumbers,floatfix]{revtex4-1}

\usepackage{dynlearn}
\usepackage{xcolor}

\definecolor{mygray}{gray}{0.6}
\newcommand{\deBrog}{\lambda_\text{d}}
\DeclareMathOperator{\Tr}{Tr}

\theoremstyle{plain}    

\begin{document}

\def\ourTitle{Szilard's Engine as a\\
Quantum Thermodynamical System}

\def\ourAbstract{We analyze an engine whose working fluid consists of a single quantum particle,
paralleling Szilard's construction of a classical single-particle engine.
Following his resolution of Maxwell's Second Law paradox using the latter,
which turned on physically instantiating the demon (control subsystem), the
quantum engine's design mirrors the classically-chaotic Szilard Map that
operates a thermodynamic cycle of measurement, thermal-energy extraction, and
memory reset. Focusing on the thermodynamic costs to observe and control the
particle and comparing these in the quantum and classical limits, we detail the
thermodynamic tradeoffs behind Landauer's Principle for
information-processing-induced thermodynamic dissipation in the quantum and
classical regimes. In particular, and as found with the classical engine, we
show that the sum of the thermodynamic costs over a cycle obeys a generalized
Landauer Principle, exactly balancing energy extraction from the heat bath.
Thus, the quantum engine obeys the Second Law. However, the quantum engine does
so via substantially different mechanisms: classically measurement and erasure
determine the thermodynamics, while in the quantum implementation the cost of
partition insertion is key.
}

\def\ourKeywords{Landauer Principle, heat engine, information engine, Maxwell Demon, quantum
thermodynamics
}

\hypersetup{
  pdfauthor={James P. Crutchfield},
  pdftitle={\ourTitle},
  pdfsubject={\ourAbstract},
  pdfkeywords={\ourKeywords},
  pdfproducer={},
  pdfcreator={}
}

\author{Maryam Ashrafi}
\email{ma.ashrafi91@gmail.com}
\affiliation{Complexity Sciences Center and Physics Department,
University of California at Davis, One Shields Avenue, Davis, CA 95616}

\author{Kyle J. Ray}
\email{kjray@ucdavis.edu}
\affiliation{Complexity Sciences Center and Physics Department,
University of California at Davis, One Shields Avenue, Davis, CA 95616}

\author{Fabio Anza}
\email{fanza@uw.edu}
\affiliation{InQubator for Quantum Simulation (IQuS), Department of Physics,
University of Washington, Seattle, WA 98195, USA}
\affiliation{Complexity Sciences Center and Physics Department,
University of California at Davis, One Shields Avenue, Davis, CA 95616}

\author{James P. Crutchfield}
\email{chaos@ucdavis.edu}
\affiliation{Complexity Sciences Center and Physics Department,
University of California at Davis, One Shields Avenue, Davis, CA 95616}

\date{\today}
\bibliographystyle{unsrt}

\title{\ourTitle}

\begin{abstract}
\ourAbstract
\end{abstract}

\keywords{\ourKeywords}

\preprint{\arxiv{2010.14652}}

\title{\ourTitle}
\date{\today}
\maketitle


\setstretch{1.1}

\newcommand{\kB}{k_\text{B}}
\newcommand {\SUS}{\text{SUS}\xspace}
\newcommand{\DEM}{\text{demon}\xspace}

\section{Introduction}

Sparked by Maxwell's thought experiment on the limitations of the Second Law
\cite{Maxw88a}, modern efforts to improve our understanding of nanoscale
processes have clearly highlighted information as a thermodynamic resource
\cite{Whee82a,Whee89a,Land91a}. Starting with Szilard's insights
\cite{Szil29a}, then going through later work by Landauer \cite{Land87a},
Bennett \cite{Benn82}, and many others \cite{Leff02a}, the modern developments
of stochastic \cite{Seif12a} and information \cite{Parr15a} thermodynamics
allow us to appreciate Maxwell's demon as an \emph{information engine}
\cite{Boyd14b,Boyd15a}---a physical system whose dynamical evolution
simultaneously stores and processes information in the service of a desired
cycle of thermodynamic transformations.

While originally formulated in the classical domain, the inherent microscopic
character of these analyses, together with dramatic improvements in our ability
to manipulate systems at the nanoscale, invites us to re-examine these engines
to account for quantum behavior. In fact, quantum Maxwell demons and
Szilard engines have been widely discussed
\cite{Mohammady17,Zurek18,Alicki19,Saga09,Saga11a,Li12a,Aydi20a,Kim11b, Cai12,
Jeon16, Beng18a, Song19, Zhua14, Lu12, Kim12a, Beng18b,Dong11} and
experimentally realized \cite{Kosk14a,Cama16a,Cama16a, Wang18a, Pete20a,Vidr16,
Cott17a, Nagi18a}. The following takes a complementary approach with respect to
existing explorations, though, as it builds on the mechanistically-detailed
version of Szilard's engine introduced in Refs. \cite{Boyd14b,Ray20a}. 

While a fully accurate description of engine operation requires accounting for
the underlying dynamics, we focus on two crucial aspects of a
quantum Szilard engine. First, recognizing the physical role of the information
processing exerted by the control subsystem, we give a microscopically-detailed
treatment of this so-called demon by giving it physical form. Second, we improve on the description of the
costs arising in each of the distinct stages of the engine's thermodynamic
cycle---measurement, thermal-energy extraction, and reset. This involves
analyzing how each stage's operation depends on the physical parameters at
play.
\begin{figure*}[t]
\centering
\includegraphics[width=\linewidth]{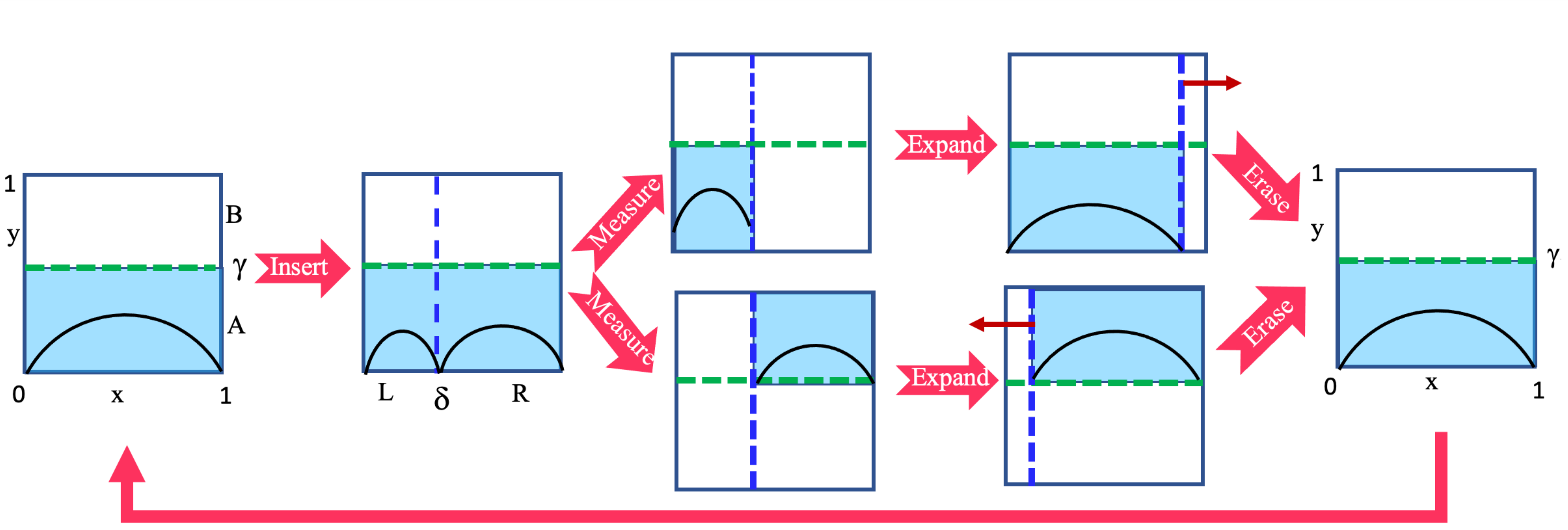}
\caption{Quantum Szilard engine thermodynamic cycle: Insert partition, measure
	particle location, expand, and erase memory of particle
	location to reset joint system state to begin next cycle. Cf. actions of
	the classical engine under control of the Szilard Map in Ref.
	\cite{Boyd14b}'s Fig. 1. The horizontal dimension ($x$) is the state of the
	quantum particle. The vertical dimension ($y$) is the state of the demon
	(control) subsystem. $\delta$ determines the location of the partition upon
	insertion (vertical dashed blue line). $\gamma$ delineates the boundary of
	the demon's two memory states (horizontal dashed green line). The solid
	blue curve is a schematic representation of the quantum particle's
	wavefunction. The light blue shaded region tracks the region of positive
	probability for the joint (particle,demon) state. 
	}
\label{fig:QSEOperation}
\end{figure*}

To do so, we parallel the dynamical-systems analysis performed in Ref.
\cite{Boyd14b} that highlighted the interplay between the entropic
cost of erasure and measurement, and how they depend on both system and demon
parameters. The result is a plethora of thermodynamic phenomena in which
entropic costs of measurement, erasure, and control are traded-off against each
other, while still respecting Landauer's Principle \cite{Land61a}.

In particular, unlike the classical engine, the quantum version incurs a cost
when inserting a partition divider. The upshot is that the quantum engine, too,
obeys the Second Law, via a generalized Landauer's Principle, though due to
different mechanisms. This achieves a more detailed exploration of the engine
as a \emph{thermodynamical system}---a dynamical system whose evolution of
microstate distributions supports macroscopic thermodynamic transformations.

Our aim is to go beyond Landauer's Principle, though, to uncover the dynamical
interplay between thermodynamic costs, dynamic evolution, and (quantum)
information processing. This provides a comprehensive picture. The
thermodynamic costs are analyzed in detail in classical and quantum regimes.
Then, before drawing conclusions, we briefly review related results on
alternative quantum Szilard engines.

\section{Engine Design}

The \emph{Quantum Szilard Engine} (QSE) is an ideal system with which to
examine the role of information processing during thermodynamic
transformations. The engine consists of two components: the \emph{system under
study} (\SUS), and a quantum controller or \emph{\DEM}. Together, they are
surrounded by an incoherent \emph{environment} that maintains the composite
system in thermal equilibrium at inverse temperature $\beta$. 

As noted above, a wide diversity of physical instantiations have been proposed.
Here, following the classical engine introduced by Ref. \cite{Boyd14b}, we
model the joint system as a single quantum particle in a 2D box of square
geometry, with sides of unit length. Horizontal and vertical axes, $x,y \in
[0,1]$, represent the continuous degrees of freedom of \SUS and \DEM,
respectively. Figure \ref{fig:QSEOperation} lays out the individual
transformation the comprise the engine's cyclic operation.

Paralleling the classical engine's thermodynamic cycle, the QSE steps through a
repeating sequence of four functionally-distinct operations or stages:
\emph{insertion}, \emph{measurement}, \emph{control}, and \emph{erasure}. Each
of these stages is implemented by changing the potential energy surface
according to a deterministic protocol.

In the first stage of the engine's cycle a potential barrier is inserted at $x
= \delta$. Then, a coarse-grained projective measurement is performed on the
\SUS position, correlating whether it is left ($L$) or right ($R$) of the
barrier with the state of the \DEM. To track the measurement outcome two
relevant informational demon states $A$ and $B$ are identified: $A$ with the
box's lower area $y \leq \gamma$ and $B$ with the upper area $y>\gamma$.
$\gamma$ controls measurement fidelity---how much particle positional
information is stored in the demon's memory states.

The result is that the joint state space (unit square) is partitioned in four
macrostates: $\lbrace AL,BL, AR,BR \rbrace$, where $\lbrace L \sim x \in
(0,\delta], R \sim x \in (\delta ,1)\rbrace$ and $\lbrace A \sim y \in
(0,\gamma], B \sim y \in (\gamma, 1) \rbrace$. The projector onto a given
macrostate's subspace can be directly written as:
\begin{align*}
\Pi_{AL} = \int_L dx \int_A dy \ket{x,y}\bra{x,y}
  ~
\end{align*}
and similarly for $\Pi_{AR}$, $\Pi_{BL}$, and $\Pi_{BR}$. Subscripts denote
informational mesostates.

Without loss of generality, we take the demon's reference (initial) state to be
$A$. Thus, if the particle is found in macrostate $L$, the \DEM need not update
its memory state; it does nothing. If the particle is found in position
mesostate $R$, though, the demon updates its memory state to mesostate $B$.
The informational mesostates of the joint system resulting from this action is
described by a CNOT gate \cite{Niel11a}, correlating the left-right measurement
outcome with the demon's upper-lower informational mesostate:
\begin{align*}
A \otimes L & \rightarrow A \otimes L \\
A \otimes R & \rightarrow B \otimes R
~.
\end{align*}

Then, the barrier is released and allowed to move (right or left) contingent on
the \DEM memory state ($A$ or $B$), respectively. In this way, energy is
extracted from the heat bath by the support of the wave-function expanding to
fill the whole volume. Finally, the composite system is reset to its original
state $A$ to prepare the engine for a new cycle. Figure \ref{fig:QSEOperation}
also shows the joint system's informational states. 

There are two relevant features to call out. First, the \SUS and \DEM operate
as a joint quantum system. Second, $\delta$ controls the barrier's location and
so the \SUS mesostate. Similarly, $\gamma$ controls how the position mesostate
maps into the \DEM informational mesostates. With this setup, the thermodynamic
analysis can explore the interplay between the resource costs and information
processing during the cycle stages as a function of $\delta$ and $\gamma$,
respectively.

The following first recalls the thermodynamics of a particle in a 1D box in the
semi-classical and quantum regimes. It then describes the quantum
thermodynamics at each engine stage. Details of the numerical evaluations and
analytical calculations will appear elsewhere.

\section{Thermodynamics of a particle in a box}

To put the analysis in context, it is helpful to first review the core aspects
of the thermodynamics of a quantum particle in a 1D box of length $\ell$. Its
energy levels are:
\begin{align*}
E_n = \frac{n^2\pi ^2 \hslash ^2 }{2m\ell^2}
  ~, \qquad n \in \mathbb{N}_+~.
\end{align*}
The partition function:
\begin{align*}
Z = \sum_n e^{-\beta E_n}
\end{align*}
can be written in terms of the Jacobi function:
\begin{align*}
Z = \frac{1}{2} \left( \theta_3(0,q)-1 \right)
  ~,
\end{align*}
where:
\begin{align*}
\theta_3(0,q) = \sum_{i=-\infty }^{\infty}q^{i^2}
  ~,
\end{align*}
with:
\begin{align*}
q \equiv \exp{\left[-\pi \left( \deBrog / 2 \ell \right)^2\right]}
  ~.
\end{align*}
 
Here, $\deBrog(T) = \sqrt{2\pi\hbar^2 / m \kB T}$ is the thermal de Broglie
wavelength, and $-\ln q=\beta E_1$ is the measure of coldness in the ground 
state energy scale. The relevant features that determine the operating regimes are (i)
the size of the box $\ell$, to be compared with $\deBrog$, and (ii) the
boundary conditions, given using the standard Dirichlet prescription of zero
value at the boundary and outside the box.

Conveniently, the wavelength $\deBrog$ (directly) and parameter $q$ (implicitly)
determine the system's degree of quantumness. This then identifies the physical
regimes:
\begin{itemize}
      \setlength{\topsep}{-5pt}
      \setlength{\itemsep}{-5pt}
      \setlength{\parsep}{-5pt}
\item Classical---$\deBrog \ll \ell$ with partition function:
\begin{align*}
Z^\text{C} = \tfrac{1}{2} \sqrt{\pi / |\ln{q}|}
  ~.
\end{align*}
\item Quantum---$\deBrog > \ell$ where we must use the exact form:
\begin{align*}
Z^\text{Q}= \left( \theta_3(0,q)-1 \right) / 2
  ~.
\end{align*}
\end{itemize}

As all thermodynamic quantities are computed via derivatives of the partition
function, a quantitative analysis that holds across all temperature needs an
accurate evaluation of the partition function and its low-degree derivatives
at any temperature.

To this end, we compare the classical approximations against a high-precision
numerical evaluation of the full partition function. The crucial difference
with the exact case is given by the change in $Z(T)$'s convexity at low
temperatures, which is not reproduced by the classical approximation. Since
$Z$'s convexity, determined by the behavior of second-order derivatives,
governs fluctuations, the classical approximation predicts dramatically
different thermal fluctuations at low temperatures. Thus, even though the
system consists of a single particle, and so lacks phase-transition-like
phenomena (e.g., Bose-Einstein condensation), a fully quantum expression is
still needed to appropriately model macroscopically relevant quantities,
together with their fluctuations, at low temperatures.

Below a temperature $T_1$ with $k_B T_1 = E_2-E_1$---i.e., $\deBrog \gg \ell$
$(q<.2)$---the system is effectively in its ground state, with a nonvanishing
energy given by the zero point and a vanishing von Neumann entropy. This means
that energy and entropy fluctuations are actively suppressed below a certain
temperature.

Assuming that at the end of each engine stage the particle subsystem
is in a Gibbs' canonical state, the internal energy is:
\begin{align*}
U & = \Tr H \rho \\
  & = - \partial \log Z / \partial \beta
\end{align*}
and the entropy is:
\begin{align*}
S & = - \kB \Tr \rho \log \rho \\
  & = \kB \left(1-\beta \tfrac{\partial }{\partial\beta }\right) \ln Z
  ~,
\end{align*}
where $\rho$ is the density matrix.

Using this framework, a sequel
reports detailed expressions for the thermodynamic bookkeeping of entropy,
internal energy, and work, as a function of the parameters $\delta$ and
$\gamma$ in three physical regimes: classical, ground state, and quantum.  It
compares the thermodynamic costs when only the ground state is populated, as is
usually done for very low temperature with the exact full quantum regime,
illustrating how increasing temperature populates additional energy levels. The
results
highlight the appreciable differences between the ground-state populated
temperature dependencies and the exact quantum dependencies that account for
higher-energy eigenstates. From this, one appreciates how both the barrier
position $\delta$ and the memory partition $\gamma$ directly affect
thermodynamic costs of insertion. Below, we highlight and discuss certain
results from this analysis. The appendix contains these results in a simplified
form.

\section{Initialization}

The QSE is prepared in a reference state in which the demon (controller
subsystem) is in a thermal state localized  to the lower part of the
box---demon mesostate $A$---while the SUS is in a thermal state with full
support on $x \in [0,1]$. Let $\rho_{x,y}(\delta,\gamma)$ be a generic Gibbs'
canonical state of the joint system and $Z_{x,y}(\delta,\gamma)$ its partition
function, where the dependence on the previously defined parameters $\delta$
and $\gamma$ is made explicit. Accordingly, we call $\rho_x(\delta)$ and
$\rho_y(\gamma)$ the Gibbs' canonical state of SUS and demon, respectively, and
$Z_x(\delta)$ and $Z_y(\gamma)$ their respective partition functions. With
these definitions, the initial configuration of the engine is $\rho_x(1)
\otimes \rho_y (\gamma)$ with partition function $Z_x(1) Z_y(\gamma)$ and
energy levels $E_{\SUS} + E_{\DEM}$. 

\section{Insertion}

Keeping in mind the classical engine's operation, we now show that the most
distinctive step in the quantum engine cycle arises from the partition's
isothermal insertion. To model insertion we follow Refs.
\cite{Joklegar09,Pedram10}'s treatment: the barrier is described by a very thin
$\delta$-function potential $V(t) = \lambda(t) \delta(x - \delta)$. As
$\lambda(t) \to \infty$, barrier insertion erases all coherence between the
particle residing on either side. This occurs for all energy eigenstates, thus
the same loss of coherence holds for canonical states.

This is a key observation that arises from a dynamical argument regarding the
partition's physical nature. So, the suppression of the off-diagonal matrix
elements occurs due to the barrier's insertion, not the projective measurement
of particle position. This step substantially modifies the thermodynamic cost
of measurement, while leaving unaltered the sum of its own cost and the
measurement's. We note that, based on a completely different argument, Ref.
\cite{Zurek18} also draws this same conclusion.

\begin{figure}[t]
\centering
\includegraphics[width=0.92\columnwidth]{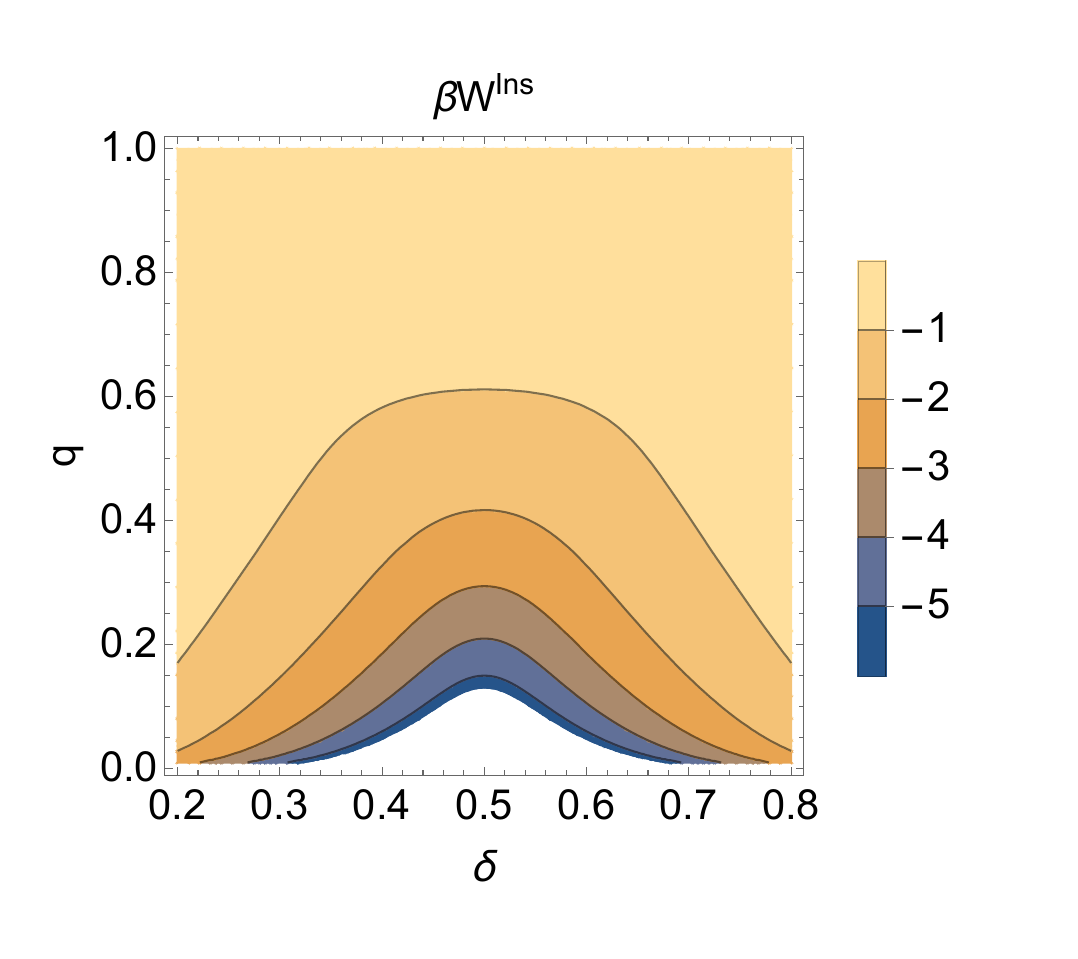}
\caption{Thermodynamic work cost $W^{\text{ins}}$ of inserting the partition
	at locations $\delta \in [0,1]$ as a function of temperature for $\hbar=1$
	and $2m=\pi^2$. The legend at right gives the values of $\beta
	W^{\text{ins}}$.  The thermodynamic work cost maximizes at $\delta = 1/2$.
	With increasing temperature the insertion work decreases, vanishing in the
	classical limit.
	}
\label{fig:QuantumInsertion}
\end{figure}

Accordingly, the energy cost due to destroying coherences does not occur due to
measurement, but instead due to insertion. This thermodynamics differs markedly
from the classical case in which insertion has no cost. That said, the
transformation renders the engine's density matrix an incoherent superposition
between left and right positions \emph{before measurement}. In this way, the
classical limit recovers the classical result of no thermodynamic cost for
insertion.

Figure \ref{fig:QuantumInsertion} shows the dependence of
insertion work as a function of partition parameter $\delta \in [0,1]$. 
At low temperature the insertion cost is very sensitive to the place 
of insertion and is maximum at $\delta = 1/2$. Increasing the temperature, 
the work cost of the insertion and its dependence on $\delta$ decreases 
as the classical case is smoothly recovered.

\section{Measurement as Induced Correlation}

In the quantum domain, the practical aspects of measurement are nontrivial.
And so, a more detailed development is necessary than for the classical engine.
The second (measurement) stage of the quantum engine's cycle comes in two
steps. First, it implements a coarse-grained projective measurement of the
\SUS. Then, it implements a CNOT gate that correlates the measurement result
($L$ or $R$) with the \DEM mesostate ($A$ or $B$).

If the particle is found on the left, the \DEM does nothing, as it was already
initialized to the $A$ mesostate. Otherwise, it changes mesostate from $A$ to
$B$. Therefore, the measurement cost contains two parts. The first is a
localization cost that is symmetric about $\delta$. The second is the cost of
storing the information in the demon's memory. It is not symmetric in $\delta$.
The asymmetry arises from the CNOT gate, which is not symmetric with respect to
$L$ and $R$.

Appropriate thermodynamic bookkeeping reveals the  average changes in
internal energy, entropy, and work.  The values are computed averaging over the
probability distribution $(p_L(\delta),p_R(1-\delta))$ of finding the particle
on the barrier's left or right side after insertion, where $p_L(\delta) = Z_x(\delta) /
\left(Z_x(\delta) + Z_x(1-\delta)\right)$ and $p_R(1-\delta) = 1-p_L(\delta)$.

\begin{figure}[t]
\centering
\includegraphics[width=.9\columnwidth]{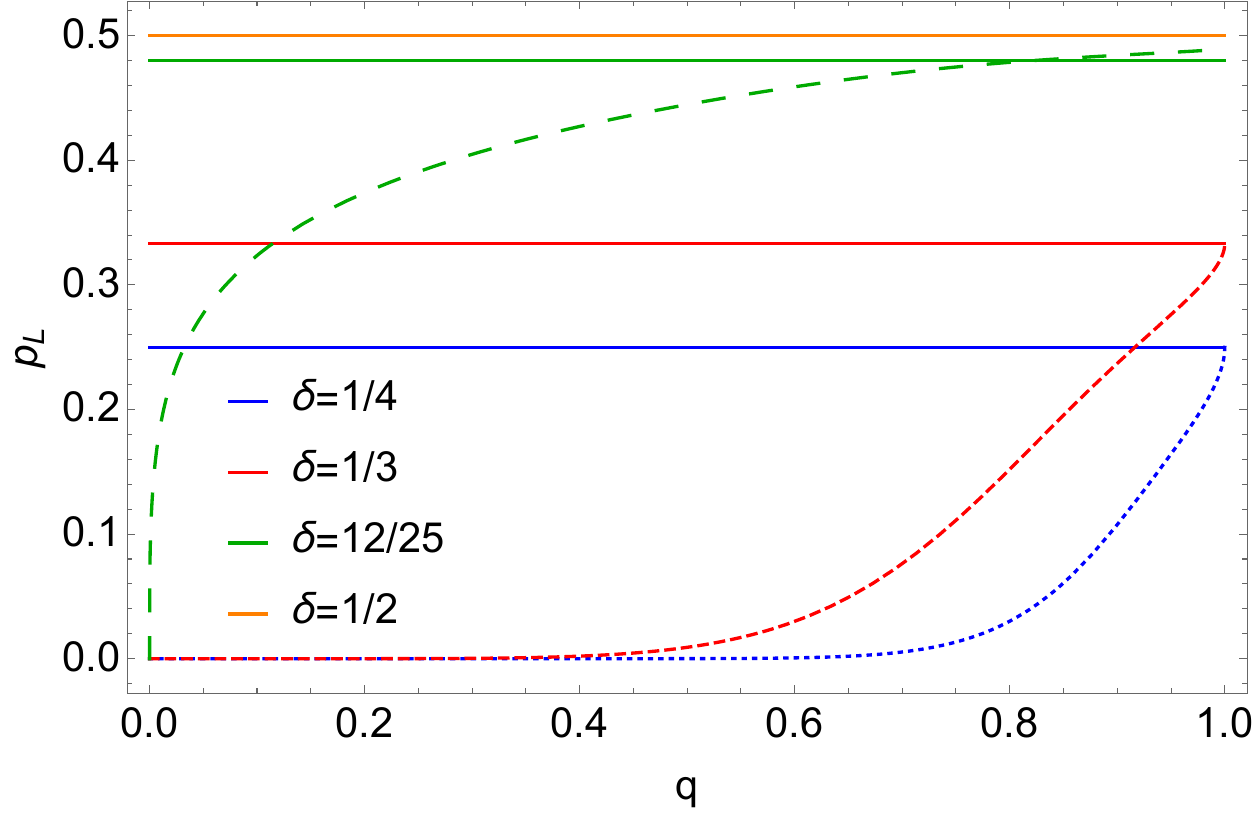}
\caption{Temperature dependence of the probability of finding the particle
	on the left side at various insertion positions $\delta \in \{1/4, 1/3,
	12/25, 1/2\}$. Dashed lines give the quantum probability and solid lines,
	classical. The validity of the classical assumption, and the ground state
	assumption are clearly affected by the interplay between the system's
	parameter $\delta$ and the temperature.
	}
\label{fig:ProbLambda}
\end{figure}

Figure \ref{fig:ProbLambda} displays $p_L$, the probability of being in the
left compartment after insertion. Classically, the probability should just be
$\delta$, the size of the partition. In the case that the system stays in its
ground state through the insertion, the particle is found with probability 1 to
be in the larger compartment, so $p_L =0$. Between these two extremes lies an
interesting regime with nontrivial behavior where the probabilities depart from
the classical case very rapidly when $q<1$ unless $\delta = .5$. Due to this,
a full quantum treatment is necessary for large swaths of the parameter space.

Moreover, the protocol's specific properties affect the thermodynamic
bookkeeping in a nontrivial way. For example, symmetric barrier insertion
($\delta = 1/2$) leads to an insensitivity to quantum effects. Sensitivity is
regained, though, for even small changes: e.g., $\delta = 12/25$. The existence
of this and other nontrivial interplay between system parameters and the
temperature highlights the need for detailed analysis of this stage.

For example, Fig. \ref{fig:ProbLambda} demonstrates that using the classical
approximation has a qualitative impact on thermodynamic accounting as the
functional dependence on the system parameters, such as temperature, are
strongly affected by quantum effects except in special cases.

\begin{figure}[ht]
\centering
\includegraphics[width=\columnwidth]{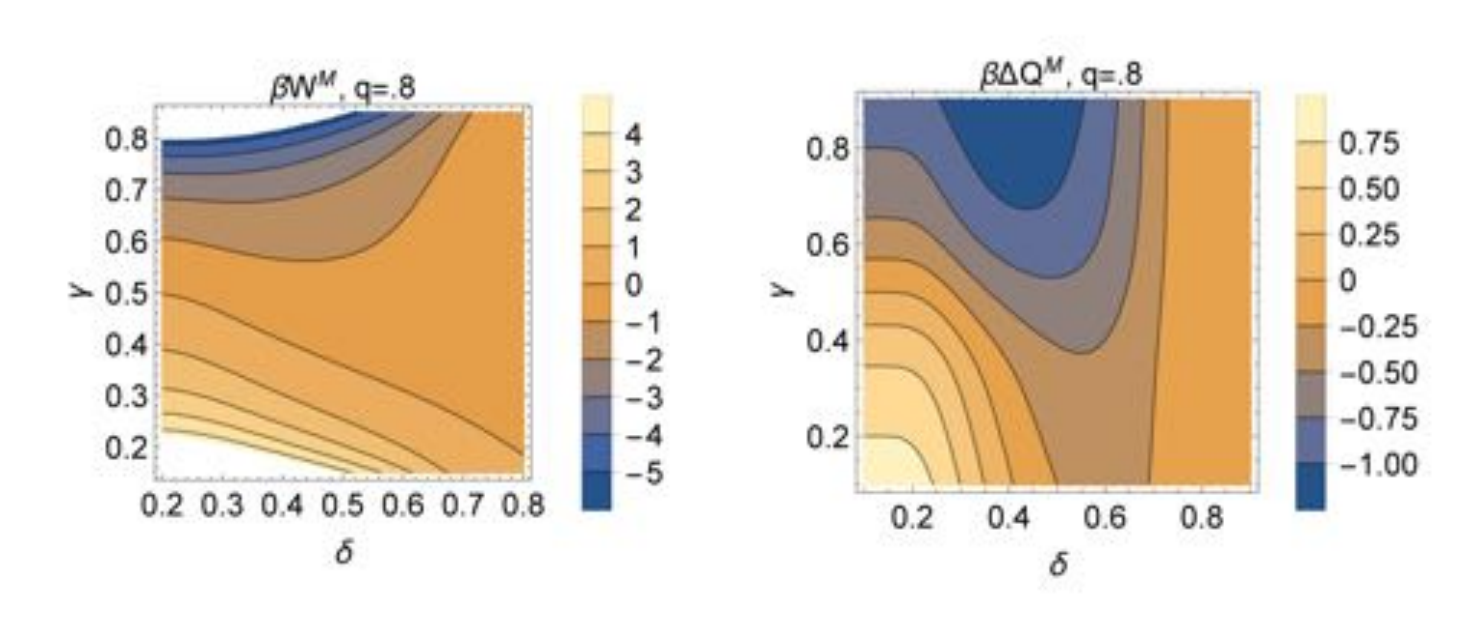}
\caption{Partition location $\delta$ and demon memory $\gamma$ dependence of
	thermodynamic heat $\beta \Delta Q_{\text{M}}$ and thermodynamic work
	$\beta \Delta W_{\text{M}}$ during measurement and correlation steps as a
	function of $\delta$ and $\gamma$ at $q=0.8$. Due to asymmetry in the
	initial memory state the measurement cost is not symmetric respect to
	$\delta$.
	}
\label{fig:Measure}
\end{figure}

To this end, Fig. \ref{fig:Measure} shows how the heat and work depend on both
the partition insertion parameter $\delta$ and demon memory parameter $\gamma$;
see App. \ref{Appendix}. Considering both the \DEM and its quantum nature allows
probing beyond $\gamma = 0.5$ to account, for example, for experimental
uncertainties. This markedly changes the thermodynamics of measurement,
resulting in nonzero contributions for both the energetic and work cost.
Furthermore, one can appreciate how the symmetry $\delta \leftrightarrow
1-\delta$---that one naively expects to hold---is present if and only if the
\DEM informational mesostates are truly equivalent, as it does not hold unless
$\gamma = 0.5$.

\section{Control}

\begin{figure}[t]
\centering
\includegraphics[width=\columnwidth]{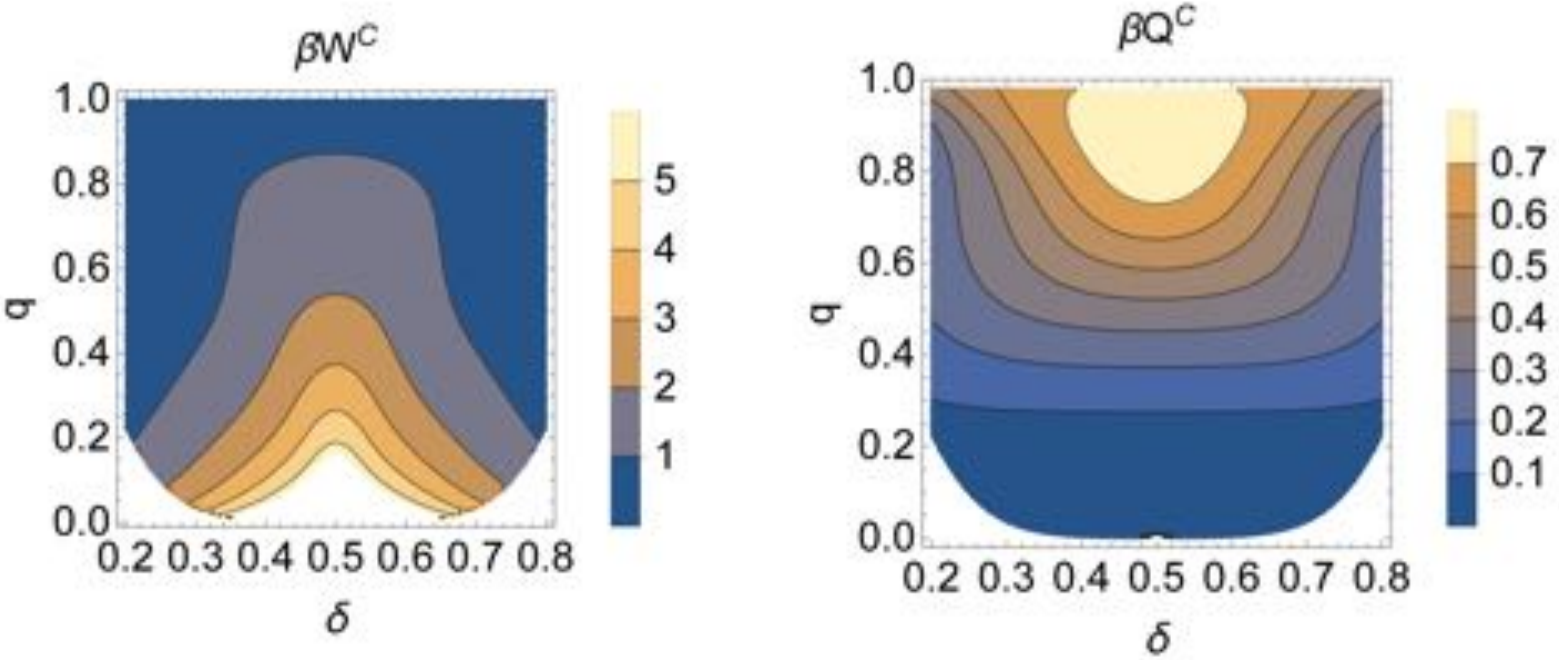}
\caption{Thermodynamic work cost $\beta W^{\text{C}}$ and heat transfer
	$\beta Q^{\text{C}}$of expansion at locations $\delta \in [0,1]$ as a
	function of $q$. For $\delta = 1/2$ the maximum work is extracted from the
	system. In the very low temperature limit, the heat transfer vanishes and
	the work done by the system equals the change of the particle's internal
	energy. At high temperatures, the internal energy change vanishes and the
	work done on the system equals the heat transfer.
	}
\label{fig:Control}
\end{figure}

After measurement and contingent on the appropriate mesostate, the partition
moves isothermally to the left (right), thereby extracting $W^{C} = -\sum_n P_n
dE_n$ of work from the ambient heat bath. In short, the control stage's
thermodynamic costs are independent of the details of the demon's informational
mesostates, while still exhibiting dependence on temperature and partition
parameter $\delta$. Figure \ref{fig:Control} reveals several interesting
results. While both the extracted work and the entropic cost monotonically
decrease as $\left\vert \delta - 0.5\right\vert$ increases, the internal energy
change depends only on the temperature and not on $\delta$. (See App.
\ref{Appendix} for the details.)

\section{Erasure}

The engine's last stage returns the joint particle-demon system to its original
configuration, preparing the engine to begin a new thermodynamic cycle.
Landauer originally claimed that all thermodynamic cost arose from the erasure
stage and not measurement \cite{Land61a}. Much later, it was shown that this
was far too restrictive \cite{Boyd14b}: depending on partition insertion and
demon memory mesostates, work and heat costs can be traded-off. As we now
show, this holds in the quantum engine, except via notably different
mechanisms.

\begin{figure}[t]
\centering
\includegraphics[width=.52\columnwidth]{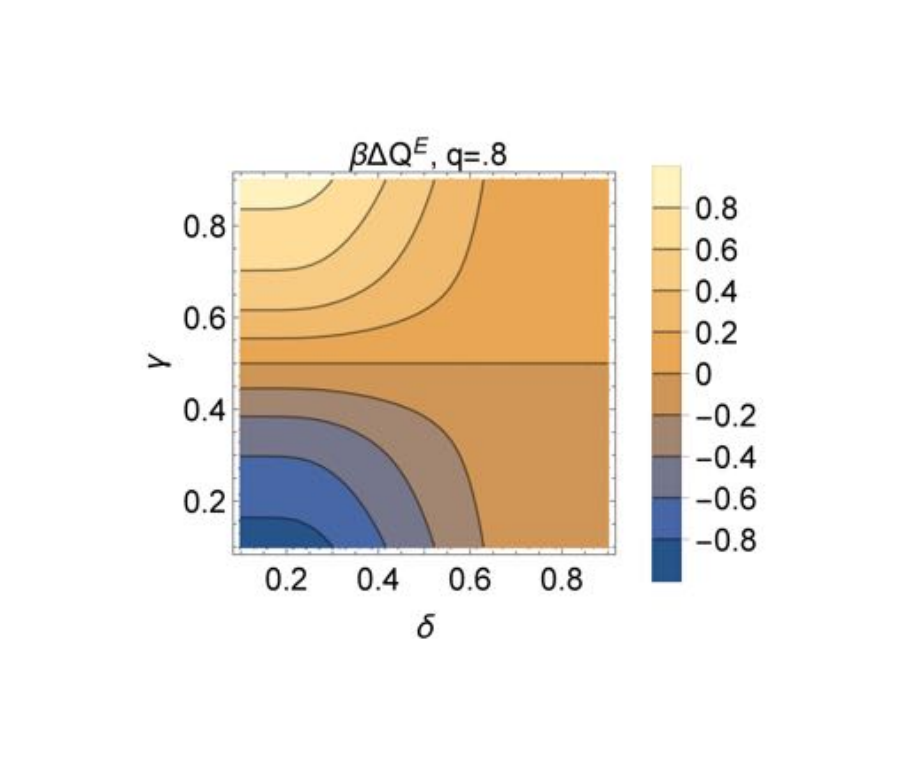}
\caption{Dissipated heat $\beta \Delta Q_E$ during erasure at $q = 0.53$.
	Nonmonotonic behavior of $\beta \Delta Q_E$ in $\left\vert \delta -
	0.5\right\vert $ results from the lack of symmetry between the
	informational mesostates at $\gamma \neq 1/2$. 
	}
\label{fig:Erase}
\end{figure}

Figure \ref{fig:Erase} reveals interesting nonmonotonic behavior in the
dissipated heat as a function of $\delta$ and $\gamma$. Here, we note that at
$\gamma = 0.5$, where the demon's informational mesostates are symmetric, there
is no work or energy cost associated with the erasure. Again, this is explained
in terms broken $\delta \leftrightarrow 1 - \delta$ symmetry---a symmetry
present only on the line $\gamma = 0.5$. While reasonable in certain cases,
this neglects structural imperfections and experimental uncertainties or even
intentionally designed differences between the demons memory states.

\begin{figure}[h]
\centering
\includegraphics[width=.9\columnwidth]{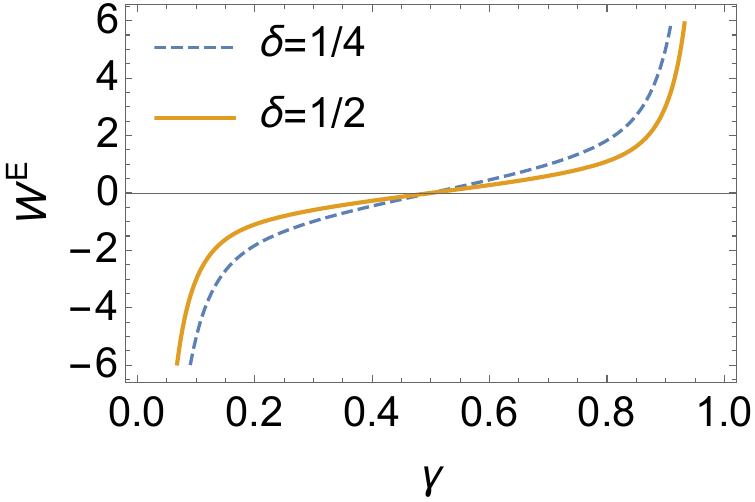}
\caption{Erasure work $W^\text{E}$ as a function of demon mesostate $A$ size
	($\gamma$) for $q=0.95$. The erasure work cost vanishes only for the 
	symmetric case $\gamma = 1/2$, even in the near-classical regime.
	}
\label{fig:ErasureWork}
\end{figure}

Figure \ref{fig:ErasureWork} confirms this by showing that the erasure work
cost vanishes only when the demon informational states have such
symmetry---viz., $\gamma = 1 -\gamma = 1/2$. If $\gamma \neq 1/2$, there is a
nonzero work-cost to account for, whose value depends on both $\gamma$ and
$\delta$.

\begin{figure}[ht]
\centering
\includegraphics[width=\columnwidth]{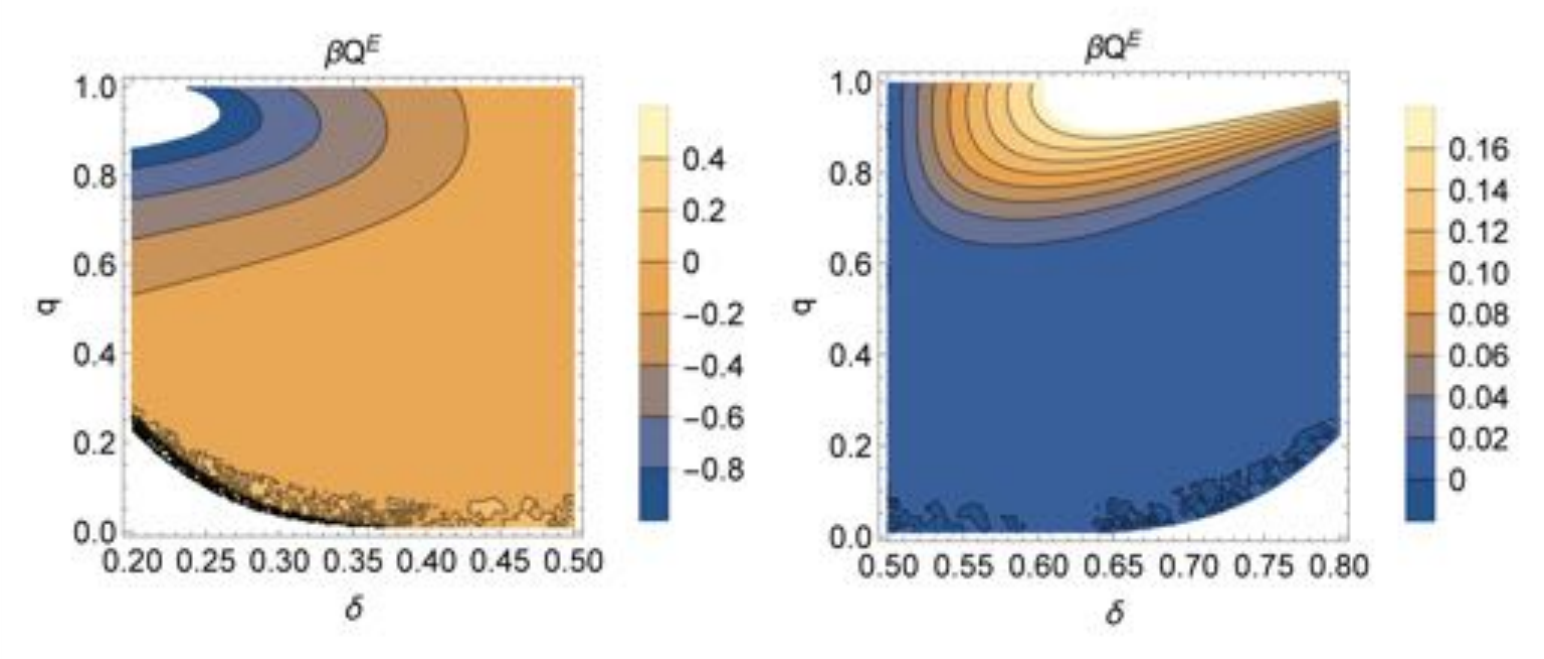}
\caption{$\beta \Delta Q_{\text{E}}$ during measurement and correlation steps
	as a function of $\delta$=$\gamma$ and q. The erasure cost vanishes for
	$\delta < .5$. This points towards the \DEM and \SUS both being in the
	ground state.	
	}
\label{fig:QEras2vsq}
\end{figure}

Meanwhile, Fig. \ref{fig:QEras2vsq} displays how quickly the ground state
can be reached when $\gamma=\delta$. Even for values of $q$ as high as $0.5$ or
$0.6$, the erasure costs can be neglected as the system settles into its
(deterministic) ground state where $p_L =0$.

\section{Discussion}

This development extends the recent dynamical analysis \cite{Boyd14b} of
Szilard's classical single-particle engine. It modeled the Szilard engine as a
quantum information engine consisting of a simple quantum system interacting
with a control system---the demon. Taking to heart Szilard's original strategy
to solve Maxwell's paradox, we analyzed the physics of both the thermodynamic
system and control system. Agreeing with Szilard, over the entire thermodynamic
cycle the net changes---resource uptake or exhaust---balance each other out.
There is zero output work and the operation is consistent with the Second Law:
\begin{align*}
\Delta U^{I}+ \Delta U^{C} & = 0 \\
\Delta U^{M}+ \Delta U^{E} & = 0
  ~, 
\end{align*}
and:
\begin{align*}
\Delta Q^{I} + \Delta Q^{C} & = -( \Delta Q^{M}+ \Delta Q^{E})~, \\
W^{I}+ W^{C} &= -( W^{M}+ W^{E})
  ~.
\end{align*}

That said, portions of the quantum engine's cycle exhibit novel behavior and
new trade-offs. In particular, they show a nontrivial relationship to
Landauer's Principle. Specifically, the aggregate entropic cost of measurement
and erasure still satisfies it (see App. \ref{Appendix}): 
\begin{align*}
\left\langle Q_{erase}\right\rangle
  + \left\langle Q_{measure} \right\rangle=  \beta^{-1} \ln 2 H(\delta) 
  ~,
\end{align*}
where $H(\delta)$ is the binary entropy function. References
\cite{Shiz95a,Fahn96a,Bark06a,Saga12a} previously noted this for the classical
engine.

\begin{figure}[t]
\centering
  (a)
\includegraphics[width=.9\columnwidth]{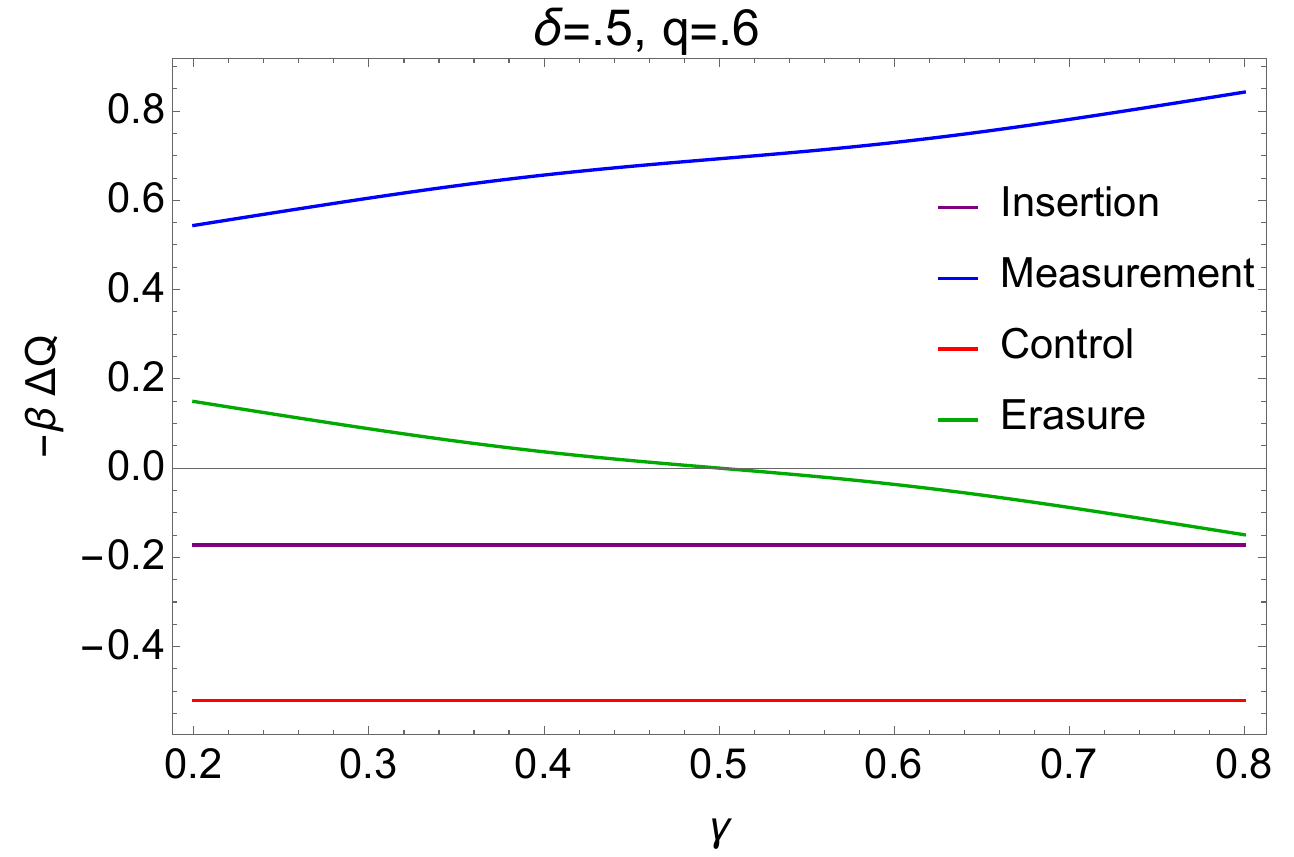}
  (b)
\includegraphics[width=.9\columnwidth]{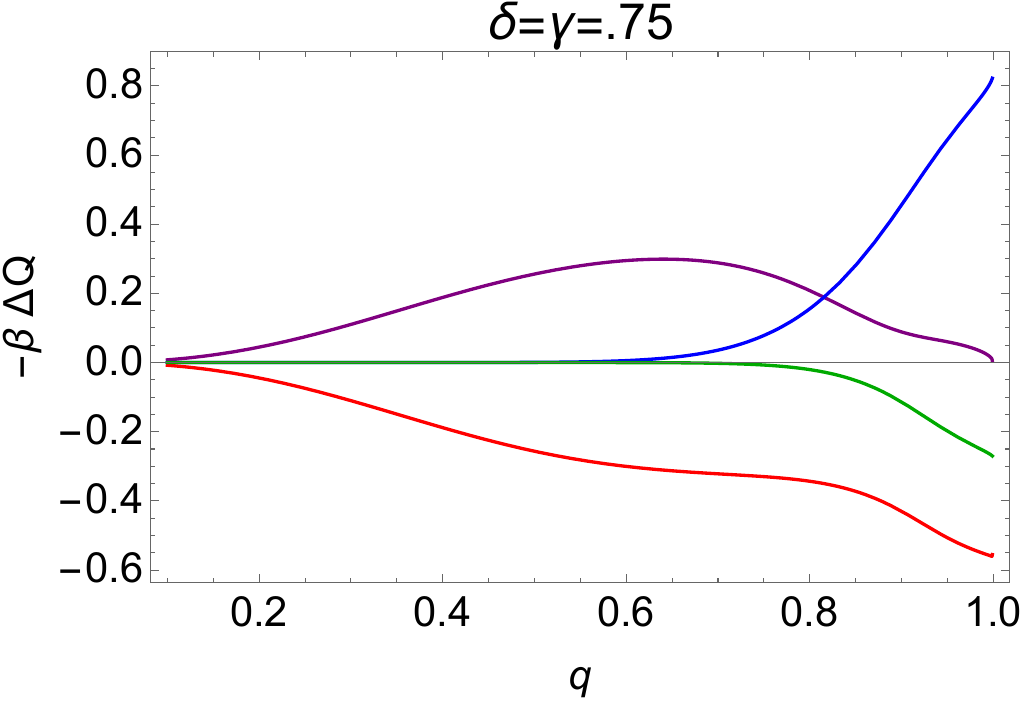} 
\caption{Thermodynamic costs $\beta \Delta Q$ versus demon memory-mesostate
	parameter $\gamma$ for measurement (blue dash line), control (red dotted
	line), and erasure (green solid line). The sum of
	measurement and erasure costs equals the sum of insertion and control
	work. (a) For $q= .6$ and $\delta = 1/2$. The cost of erasure is zero for
	$\gamma =1/2$. (b) Thermodynamic costs $\Delta Q/ \kB T$ versus $q$ for
	$\delta=\gamma=.75$. In the low temperature limit, the cost of the
	measurement and erasure vanish, while at high temperature the cost of
	insertion vanishes. Note that once in the ground state limit---here,
	approximately $q<.5$---the cost of measurement and erasure vanishes and the
	cost of insertion equals the expansion cost.
	}
\label{fig:ThermoCostsVersusLambda}
\end{figure}

While the Second Law is not violated---all entropic contributions sum to zero,
see Fig. \ref{fig:ThermoCostsVersusLambda}---the thermodynamic signature of the
individual stages varies significantly. Thus, these thermodynamic trade-offs
are key to designing quantum engines.

For example, at $\delta = \gamma = 1/2$, the entropic cost of erasure vanishes,
``violating'' Landauer's Principle. However, it still respects the trade-off
thanks to the fact that $\left\langle Q_{measure}\right\rangle = \log 2$. The
bottom panel of Fig. \ref{fig:ThermoCostsVersusLambda} illustrates the change
of the thermodynamic costs $\Delta Q/ \kB T$ from very low to very high
temperatures. In the ground state regime measurement cost and erasure cost
vanish and the insertion cost equals in magnitude the control cost. Looking at
the other end of the spectrum: in the very high temperature limit, insertion
cost vanishes and the sum of the erasure and control costs equal (in magnitude)
the measurement cost.

\section{Related work}

As the introduction noted, quantum Szilard engines have been widely analyzed
\cite{Mohammady17,Zurek18,Alicki19,Saga09,Saga11a,Li12a,Aydi20a,Kim11b, Cai12,
Jeon16, Beng18a, Song19, Zhua14, Lu12, Kim12a, Beng18b,Dong11,Shen20a} and
experimentally implemented \cite{Kosk14a,Cama16a,Cama16a, Wang18a,
Pete20a,Vidr16, Cott17a, Nagi18a}, forming an interesting set of results that
collectively aim to clarify the physical role of information processing in
nanoscale systems and its interplay with the energetics.

That said, the specific motivating questions, implementations, and
interpretations differ widely across these works. Here, we briefly mention
several with the goal of providing a bird's eye view and highlighting the
advantages of the dynamical-maps approach we pursued here. We begin surveying
theoretical results.

Zurek and Lloyd's work \cite{Zurek86,Lloyd97} was the first of the modern wave
of contributions leveraging tools from quantum information theory and quantum
thermodynamics. See also the more recent reviews of Refs.
\cite{Zurek18,Saga09,Saga11a,Aydi20a,Dong11}. Interestingly, Refs.
\cite{Kim12a,Pal21a} highlight differences between single and multi-particle
Szilard engines. While Refs. \cite{Li12a,Brac14a,Gea02a,Sord19a} show how,
despite equilibrium assumptions, the thermodynamic costs of each stage of the
quantum engine can qualitatively depend on details of the quantum model, while
recovering the classical results in the high temperature limit. 

Over the last decade or so, there also has been a number of experimental
investigations reported under the rubric of quantum Maxwell demons and Szilard
engines. In this, one finds an even wider variation on what constitutes
physically implementing demons.

To the best of our knowledge, Ref. \cite{Kosk14a} was the first experimental
implementation of a quantum Szilard engine that extracts $\kB T \ln 2$ of work
per one bit of information. Analogously, Ref. \cite{Pete20a} realized an
NMR-based quantum Szilard engine, while Ref. \cite{Cott17a} implemented a
direct ``information to work'' conversion with a superconducting (transmon)
platform. Maxwell demons have also been implemented on a photonic platform
\cite{Vidr16} and on solid-state spin systems \cite{Wang18a}.

Overall, there is agreement that these constructions are consistent with the
Second Law, once the appropriate role of information processing is recognized.
However, stage-specific thermodynamic costs can depend on model details,
complicating a universal understanding of the interplay between energetics and
information processing. 

This diversity motivated our parametrized analysis---an exploration that
accounts for model details via the parameters $\delta$ and $\gamma$. The
analysis also addressed the difference in behaviors in the classical and
quantum regimes. Moreover, reflected on this background, our analysis differs
in that it starts with a quantum counterpart of constructions in Ref.
\cite{Boyd14b,Ray20a} that explicitly implement each of Szilard's
transformations. In effect, each step in the engine's thermodynamic cycle is a
piecewise linear map on the macrostates of the joint system, with an associated
quantum transformation---essentially a particle in a 2D box with time-varying
boundaries. This led to detailed thermodynamic bookkeeping for each engine
stage, with quantitative results that track parameter dependence. Modeling the
engine stages with dynamical maps provided a direct connection to quantum
dynamical systems. The latter's perspective is often not included in quantum
thermodynamics of nanoscale systems. For that matter, the conclusion from the
classical analysis---that the physics and information dynamics are both
essential---holds with extra force in the quantum domain.

\vspace{-0.2in}
\section{Conclusion}
\vspace{-0.1in}

We revisited Szilard's engine from the broader perspective of a quantum
information engine---a quantum machine that manipulates both energy and
information to produce work while dissipating heat. Including both the
parametric dependence on the informational states of the demon, and its quantum
nature, into the engine's description allowed us to explore a variety of
nonclassical thermodynamic behaviors. Those behaviors, while compatible with
the principles of quantum thermodynamics, exhibited new and different
thermodynamic signatures. We believe these analyses generalize to arbitrary
quantum information engines and will be an aid in designing efficient quantum
information processing devices.

\vspace{-0.2in}
\section*{Acknowledgments}
\label{sec:acknowledgments}
\vspace{-0.2in}

F.A. and J.P.C. thank the Telluride Science Research Center for its hospitality
during visits and the participants of its annual Information Engines summer
workshop for stimulating discussions. This material is based on work supported
by, or in part by, a Templeton World Charity Foundation Power of Information
Fellowship, Foundational Questions Institute Grant FQXi-RFP-IPW-1902, and U.S.
Army Research Laboratory and the U. S. Army Research Office grants
W911NF-18-1-0028 and W911NF-21-1-0048.

\vspace{-0.2in}

\appendix

\section{}
\label{Appendix}
\vspace{-0.2in}

The following summarizes the analytical results derived in a sequel,
used for both plotting and discussion.

\paragraph*{Before Insertion:}
\begin{align*}
Z^{(0)} & = Z_{x}(1)Z_{y}^{A}(\gamma)
~, \\
\rho ^{(0)} & = \rho _{x}(1)\rho _{y}^{A}(\gamma)
  ~, \\
U^{(0)} & = -\dfrac{\partial }{\partial\beta }\ln Z^0=U_{x}(1)+U_y^{A}(\gamma)
~, \\
S^{(0)} & = k_{B}\left(1-\beta\dfrac{\partial }{\partial\beta }\right)\ln Z^{0}
=S_{x}(1)+S_y^{A}(\gamma)
~.
\end{align*}

\paragraph*{After Insertion:}
\begin{align*}
Z^{(I)} & =\left(Z_{x}^{L}(\delta)+Z_{x}^{R}(1-\delta)\right)Z_{y}^{A}(\gamma)
~, \\
\rho ^{(I)} & =\left(p^{L}(\delta)\rho _{x}^{L}(\delta)+p^{R}(1-\delta)\rho _{x}^{R}(1-\delta)\right)
\rho _{y}^{A}(\gamma)
~, \\
U^{(I)} & =-\dfrac{\partial }{\partial\beta }\ln Z^{I}
~, \\
S^{(I)} & =k_{B}\left(1-\beta\dfrac{\partial }{\partial\beta }\right)\ln Z^{I}
~.
\end{align*}
Here $p^{L}(\delta)$ and $ p^{R}(1-\delta)$ are the probabilities of 
finding the particle in the left and the right sides of the box,
respectively:
\begin{align*}
p^{L}(\delta) & = \dfrac{Z_{x}(\delta)}{Z_{x}(\delta)+Z_{x}(1-\delta)}
~\text{and} \\
p^{R}(1-\delta) & = \dfrac{Z_{x}(1-\delta)}{Z_{x}(\delta)+Z_{x}(1-\delta)}
 ~.
\end{align*}

\vspace{-0.2in}
\paragraph*{Measurement:}
The average change of the internal energy, heat transfer, and work after measurement are:
\begin{align*}
\langle\Delta U^{M}\rangle 
  & = -p^{R}(1-\delta)\dfrac{\partial }{\partial\beta }
\left(\ln\dfrac{Z^B(1-\gamma)}{Z^A(\gamma)}\right)
  ~, \\
\dfrac{\langle\Delta S^{M}\rangle }{k_B}
  & = p^{L}(\delta)\ln p^{L}(\delta)+p^{R}(1-\delta)
\ln p^{R}(1-\delta)\\\nonumber
  & \qquad + p^{R}(1-\delta)\left(1-\beta\dfrac{\partial }{\partial\beta }\right)
\ln\dfrac{Z^B(1-\gamma)}{Z^A(\gamma)}
~, \\
\beta\langle W^{M}\rangle
  & = p^{L}(\delta)\ln p^{L}(\delta) \\
  & \qquad + p^{R}(1-\delta)\left[\ln p^{R}(1-\delta)+\ln\dfrac{Z^B(1-\gamma)}{Z^A(\gamma)}\right]
  ~.
\end{align*}

\paragraph*{Control:}
The average change of the internal energy and the work extracted 
from the isothermal expansion are:
\begin{align*}
\langle\Delta U^{C}\rangle
  & = -\dfrac{\partial }{\partial\beta }
  \ln\frac{Z_x(1)}{ \left(Z_x(\delta) +Z_x(1-\delta)\right)}
~\text{and} \\
\beta\langle\Delta W^{C}\rangle
  & = \ln Z_x(1)-p^{L}(\delta)\ln Z_x(\delta)\\
& \qquad - p^{R}(1-\delta)\ln Z_x(1-\delta)
~.
\label{WC}
\end{align*}

\paragraph*{Erasure:}
The average change of the internal energy, entropy, and the work extracted 
during erasure are:
\begin{align*}
\langle\Delta U^{E}\rangle
  & = -p^{R}(1-\delta)\dfrac{\partial }{\partial\beta }
\ln\dfrac{Z^A(\gamma)}{Z^B(1-\gamma)}
~, \\
\langle\Delta S^{E}\rangle
  & = p^{R}(1-\delta) \kB \left(1-\beta\dfrac{\partial }
{\partial\beta }\right)\ln\dfrac{Z^A(\gamma)}{Z^B(1-\gamma)}
~, \\
W^{E} & = p^{R}(1-\delta)\beta\ln\dfrac{Z^A(\gamma)}{Z^B(1-\gamma)}
~.
\end{align*}

\end{document}